\documentclass[twocolumn]{aastex631}
\usepackage{xcolor}
\usepackage{amsmath}

\usepackage{graphicx}
\usepackage{txfonts}
\usepackage[normalem]{ulem}

\begin{document}

   \title{Impact of current uncertainties in the $^{12}$C+$^{12}$C nuclear reaction rate on intermediate-mass stars and massive white dwarfs}

\author{Francisco C. De Gerónimo}
\affiliation{Instituto de Astrof\'{\i}sica de La Plata, CONICET-UNLP, La Plata, Argentina}

\author{Marcelo M. Miller Bertolami}
\affiliation{Instituto de Astrof\'{\i}sica de La Plata, CONICET-UNLP, La Plata, Argentina}

\author{Tiara Battich}
\affiliation{Max-Planck-Institut für Astrophysik, Karl-Schwarzschild Strasse 1, 85748 Garching, Germany}

\author{Xiaodong Tang}
\affiliation{Institute of Modern Physics, Chinese Academy of Sciences, Lanzhou 730000, China}
\affiliation{School of Nuclear Science and Technology, University of Chinese Academy of Sciences, Beijing 100049, China}

\author{Márcio Catelan}
\affiliation{Instituto de Astrof\'{\i}sica, Pontificia Universidad Cat\'olica de Chile, Av. Vicuña Mackenna 4860, 7820436 Macul, Santiago,  Chile}
\affiliation{Millennium Institute of Astrophysics, Nuncio Monse\~{n}or Sotero Sanz 100, Of. 104, Providencia, Santiago, Chile }

\author{Alejandro H. Córsico}
\affiliation{Instituto de Astrof\'{\i}sica de La Plata, CONICET-UNLP, La Plata, Argentina}

\author{Yunjun Li}
\affiliation{China Institute of Atomic Energy, Beijing 102413, China}

\author{Xiao Fang}
\affiliation{Sino-French Institute of Nuclear Engineering and Technology, Sun Yat-sen University, Zhuhai 519082, China}

\author{Leandro G. Althaus}
\affiliation{Instituto de Astrof\'{\i}sica de La Plata, CONICET-UNLP, La Plata, Argentina}

\begin{abstract}

 Recent determinations of the total rate of the $^{12}$C+$^{12}$C nuclear reaction show non-negligible differences with the reference reaction rate commonly used in previous stellar simulations. In addition, the current uncertainties in determining each exit channel constitute one of the main uncertainties in shaping the inner structure of super asymptotic giant branch stars that could have a measurable impact on the properties of pulsating ultra-massive white dwarfs (WDs). 
   We explore how new determinations of the nuclear reaction rate and its branching ratios affect the evolution of WD progenitors. 
      We show that the current uncertainties in the branching ratios constitute the main uncertainty factor in determining the inner composition of ultra-massive WDs and their progenitors. We found that the use of extreme branching ratios leads to differences in the central abundances of $^{20}$Ne of at most 17\%, which are translated into differences of at most 1.3 and 0.8\% in the cooling times and size of the crystallized core. However, the impact on the pulsation properties is small, less than 1 $s$ for the asymptotic period spacing. We found that the carbon burns partially in the interior of ultra-massive WD progenitors within a particular range of masses, leaving a hybrid CONe-core composition in their cores. The evolution of these new kinds of predicted objects differs substantially from the evolution of objects with pure CO cores. Differences in the size of the crystallized core and cooling times of up to 15 and 6\%, respectively leading to distinct patterns in the period spacing distribution.
    
\end{abstract}

 
\section{Introduction}\label{sec:int}

Stars more massive than $\sim 7 \, M_\odot$ undergo carbon fusion in the core \citep{2013sse..book.....K}.\footnote{The exact mass threshold depends on metallicity and treatment of convective boundary mixing.}  
The burning of $^{12}$C occurs at temperatures above $6\times 10^8$ K through the formation of compound nuclear states of $^{24}$Mg, denoted as $^{24}$Mg$^*$,  with excitation energies of 14 to 17 MeV above the ground level. The unstable  $^{24}$Mg$^*$ states then decay through at least five channels \citep{2020ChPhC..44k5001L}, namely: 
\begin{align}
^{12}\mathrm{C} + ^{12}\mathrm{C} \, \rightarrow \, ^{24}\mathrm{Mg}^{*} \,
& \rightarrow \, ^{20}\mathrm{Ne} + \alpha && (Q=4.62 \, \mathrm{MeV}), \nonumber\\
& \rightarrow \, ^{23}\mathrm{Na} + p  && (Q=2.24 \, \mathrm{MeV}), \nonumber \\
& \rightarrow \, ^{23}\mathrm{Mg} + n  && (Q=-2.60 \, \mathrm{MeV}),\nonumber \\
& \rightarrow \, ^{16}\mathrm{O} + ^{8}\!\mathrm{Be}  && (Q=-0.20 \, \mathrm{MeV}),\nonumber \\ 
& \rightarrow \, ^{24}\mathrm{Mg} + \gamma && (Q=14.93 \, \mathrm{MeV}). 
\end{align}
At the typical energies of carbon fusion in stars, the first two channels have similar (high) probabilities, making them many orders of magnitude more likely than the other channels. Consequently, the reactions $^{12}$C$(^{12}$C,$\alpha)^{20}$Ne and $^{12}$C$(^{12}$C$,p)^{23}$Na dominate the total $^{12}$C+$^{12}$C cross section. 
The cross-section of these reactions must be known with high accuracy down to the Gamow peak energy $E_G=1.5\pm0.3$~MeV \citep[for 5$\times10^8$ K;][]{1988ccna.book.....R} since it not only affects the production of $^{20}$Ne and $^{23}$Na, but also the subsequent evolutionary stages.
The branching ratios of the $\alpha$ and $p$ exit channels determine the total amount of $^{20}$Ne and $^{23}$Na produced inside stars. Previous works adopt 56\% \citep[][hereinafter CF88]{1988ADNDT..40..283C} and 65\% \citep{2022A&A...660A..47M} as the branching ratio for the $\alpha$ channel. However, the probability of each exit channel becomes very uncertain at the typical temperatures that characterize C-burning \citep{2013ApJ...762...31P}.

Despite considerable experimental efforts, the total $^{12}$C+$^{12}$C
reaction rate remains uncertain at stellar
temperatures.
On the one hand, the heavy ion fusion studies performed by \cite{2007PhRvC..75e7604J} suggested that the fusion cross-section may be hindered at low energies, resulting in rates lower than the standard ones from CF88. On the other hand, low-energy experiments by \citet{2007PhRvL..98l2501S} hint at the presence of resonant structure effects at lower energies that are not considered in many works and would lead to an important enhancement of the  $^{12}$C+$^{12}$C fusion rate at stellar temperatures. In the last decade, a significant effort has been made by the nuclear physics community, both experimentally and theoretically, to understand the challenging regime of astrophysical low energies of the  $^{12}$ C+$^{12}$C fusion reaction \citep{2018PhRvC..97a2801J,2018PhRvC..98f4604C,2020ChPhC..44k5001L,
Mukhamedzanov2022, 
2022EPJWC.26001002T,2023EPJWC.27911005M}. As recently reported by \cite{2022A&A...660A..47M}, the present uncertainty of the $^{12}$C+$^{12}$C rate still covers orders of magnitude at the range of temperatures of interest for astrophysical applications.

Stars with initial masses in the range $7\,M_{\odot}\lesssim M_{\rm ini}\lesssim 10 \,M_{\odot}$   might eventually become massive white dwarfs (WD). While most WDs comprise He or CO cores, these massive WDs have O and Ne as their main ingredients. 
These objects are the result of the evolution of progenitor stars that reach temperatures high enough to ignite their CO cores under degenerate conditions \citep{1994ApJ...434..306G},  as they evolve into the so-called super asymptotic giant branch (SAGB) phase. 
Classic works by \cite{1994ApJ...434..306G}, \cite{ 1996ApJ...460..489R}, \cite{1997ApJ...485..765G}, and \cite{1997ApJ...489..772I} showed that the C-flash and subsequent C-burning leads to an oxygen-neon (ONe) core and, consequently, to an ONe WD \citep[see][and references therein]{2006A&A...448..717S, 2007A&A...476..893S, 2010A&A...512A..10S, 2019A&A...625A..87C} or an electron-capture supernova \citep{2013ApJ...771L..12T,2017PASA...34...56D}, depending on the intensity of winds.
 
The chemical structure of SAGB progenitors at the end of the C-burning phase, and thus at the WD stage, depends on how the C-burning proceeds. In this sense, the current uncertainties in the total reaction rate for C-burning and its branching ratios could have a non-negligible impact on the predicted structure of ultra-massive WD. The
impact of the uncertainty of the $^{12}$C+$^{12}$C fusion rate on the properties of massive stars has been studied in several works \citep{2007PhRvC..76c5802G, 2012MNRAS.420.3047B, 2013ApJ...762...31P, 2021ApJ...916...79C, 2022A&A...660A..47M, 2024arXiv240418662D}. However, none of them explored the consequences of such uncertainties on the evolution of intermediate-mass stars and the final composition of ultra-massive WDs. Additionally, the properties of the nonradial $g$-modes of pulsating WDs depend also on the inner distribution of elements \citep[see, for example,][]{ 2017A&A...599A..21D,2022A&A...659A.150D,2021A&A...646A..30A}. In this regard, the internal chemical profile left at the end of the C-burning phase plays an important role in the pulsation properties of ultra-massive pulsating WDs \citep[see][and references therein]{2019A&ARv..27....7C, 2019A&A...621A.100D}.  
  
 In this paper, we explore how the new measurements of the $^{12}$C+$^{12}$C nuclear reaction rate and the branching ratio adopted during C-burning affect the final chemical composition and pulsations of ultra-massive WDs. The paper is organized as follows: in Section \ref{sect:uncert}, we discuss the current status of the uncertainties in both the nuclear reaction rate and its branching ratios. In Section \ref{sect:models}, we introduce the most important features of the computation of our numerical models, while in Sections \ref{sect:impact-ste} and \ref{sect:consequence} we explore the impact of those uncertainties on the properties of SAGB progenitors and pulsating ultra-massive WDs.  Finally, in Section~\ref{sect:concl}, we provide some concluding remarks.
 
\section{Uncertainties in the nuclear reaction rate and branching ratios}
\label{sect:uncert}

CF88 set a milestone in developing analytical formulae for several nuclear reaction rates, later used as standards in the computation of astrophysical numerical models. The $^{12}$C+$^{12}$C nuclear reaction stands out as very important for stellar evolution and yet remains subject to large uncertainties.
This reaction has been studied intensively in recent decades \citep[see][and references therein] {1969ApJ...157..367P,1981ZPhyA.303..305B,2007PhRvL..98l2501S,2015PhRvL.114y1102B}.  Despite the experimental efforts made and the consistent results obtained for energies near the Coulomb barrier, the total $^{12}$C+$^{12}$C fusion reaction rate remains uncertain at temperatures of astrophysical interest, with different experiments 
differing substantially \citep[see][and references therein for a detailed discussion]{2013ApJ...762...31P,2020ChPhC..44k5001L,2022A&A...660A..47M}. This is due both to the fact that the experimental background noise is high at low energies, as well as to the fact that extrapolation of the experimental data to energies below the Coulomb barrier is affected by the strong resonant behavior of the $^{12}$C+$^{12}$C cross-section. Recent studies suggest that the latter may suffer from hindrance phenomena at low energies \citep{2007PhRvC..75a5803J}, resulting in a lower rate than the widely used one from CF88. Additionally, resonant structures that would increase the reaction rate have been found at low energies \citep{2007PhRvL..98l2501S}.



\begin{figure}
  \includegraphics[width=1\columnwidth]{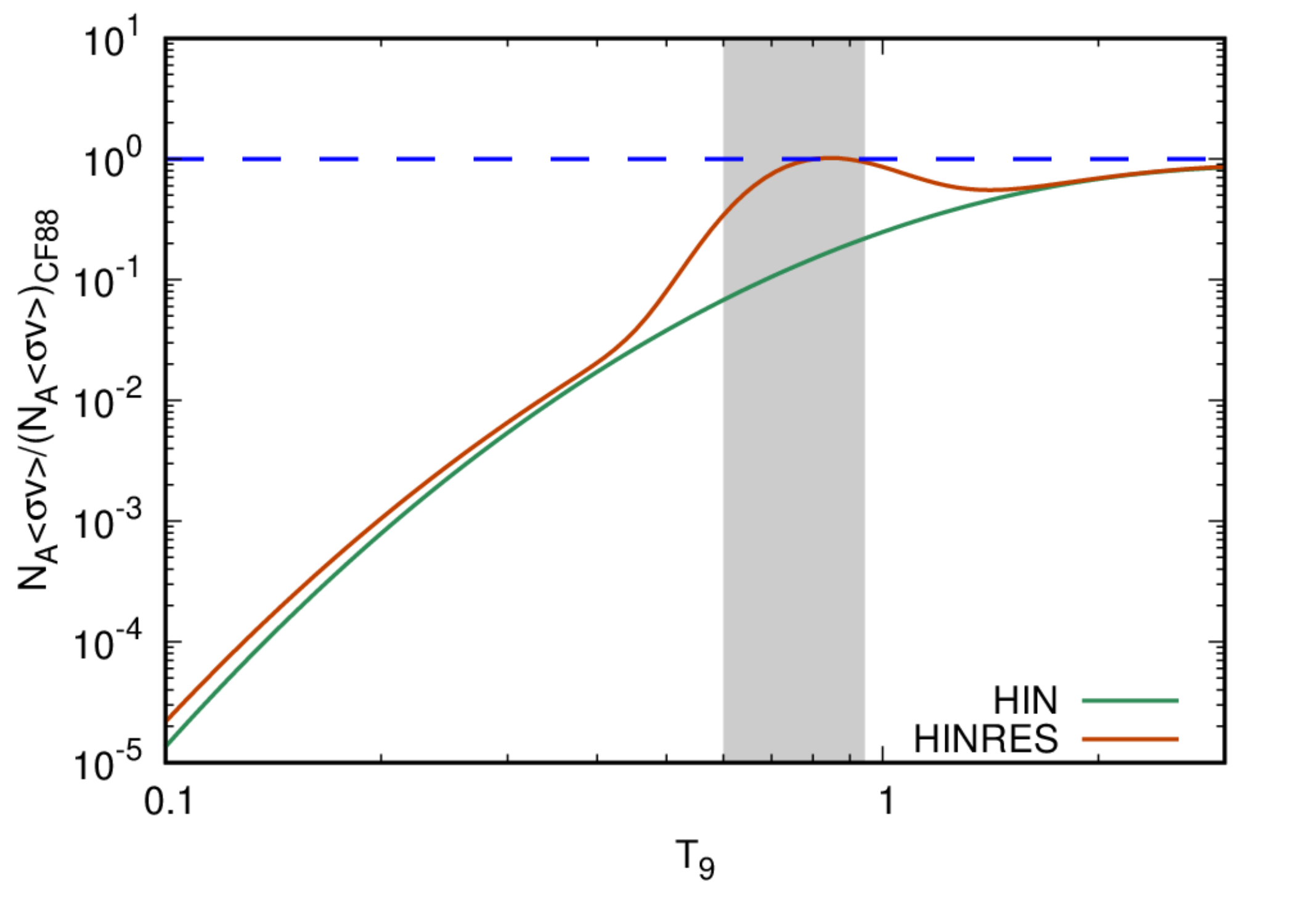}
  \caption{Ratio of the total carbon burning reaction rate between \cite{2022A&A...660A..47M} and CF88 as a function of the temperature $T_9=T/10^9$. The green line represents the HIN rates, whereas the HINRES rates are shown in red. The dashed blue line represents a ratio of 1. The shaded region depicts the range of typical temperatures characterizing $^{12}$C burning inside SAGB stars.}
  \label{fig:rate}
\end{figure}

Recently, \cite{2022A&A...660A..47M} derived the most up-to-date reaction rates for carbon fusion, based on the measurement published by the STaged ELectron Laser Acceleration (STELLA) experiment \citep{2020PhRvL.124s2701F}. In that work, the authors provided two updated formulae that account for the fusion hindrance phenomenon,\footnote{The fusion hindrance phenomenon corresponds to a sudden fall-off of the cross-section.} labeled HIN and HINRES in their study, based on different assumptions for the resonant behavior at low energies.  Specifically, the HINRES rate differs from the HIN one in that the former includes the effect of a possible low-energy resonance \citep{2007PhRvL..98l2501S}, better fitting the measured cross-section. These formulae can be incorporated into the computation of stellar models.
In Fig. \ref{fig:rate} we show the ratio between the total HIN (green line) and HINRES (red line) reaction rates and the standard value from CF88 as a function of the temperature $T_9$ (in units of $10^9$K). For the typical temperatures characterizing C-burning regions inside stars (shaded region), the HINRES rate behaves very similarly to the one provided by CF88, whereas the difference with respect to the HIN rate can reach more than a factor of ten. 

In addition to the uncertainties in the total $^{12}$C+$^{12}$C$\,\longrightarrow\,^{24}$Mg$^*$ cross-section, there are large uncertainties in the $^{24}$Mg$^*$ decay channels. The uncertainty in the relative strengths of the $^{24}$Mg$^*$ $\alpha$- and $p$-decay channels has a relevant impact on stellar nucleosynthesis calculations. The original branching ratio provided by CF88 was [56/44] for the [$\alpha/p$] channels, respectively. \cite{2013ApJ...762...31P} adopted branching ratios of [65/35] instead, and explored the consequences of extreme values, [5/95] and [95/5] using a simple single-zone post-processing calculation. Recently, a Chinese-lead international collaboration has been reanalyzing the $^{12}$C+$^{12}$C reaction \citep{2020PhLB..80135170Z,2020PhLB..80335278Z}, and, in particular, its branching ratios \citep{2020ChPhC..44k5001L}. 
Preliminary results indicate that, in the range of relevant stellar temperatures ($0.6<T_9<0.9$), the relative strength of the $\alpha$-channel to the $p$-channel decreases by about 30\% with temperature. More importantly, the uncertainty in the relative strengths of both decay channels encompasses one order of magnitude (see next sections). 


     \begin{figure}
  \includegraphics[width=1.\columnwidth]{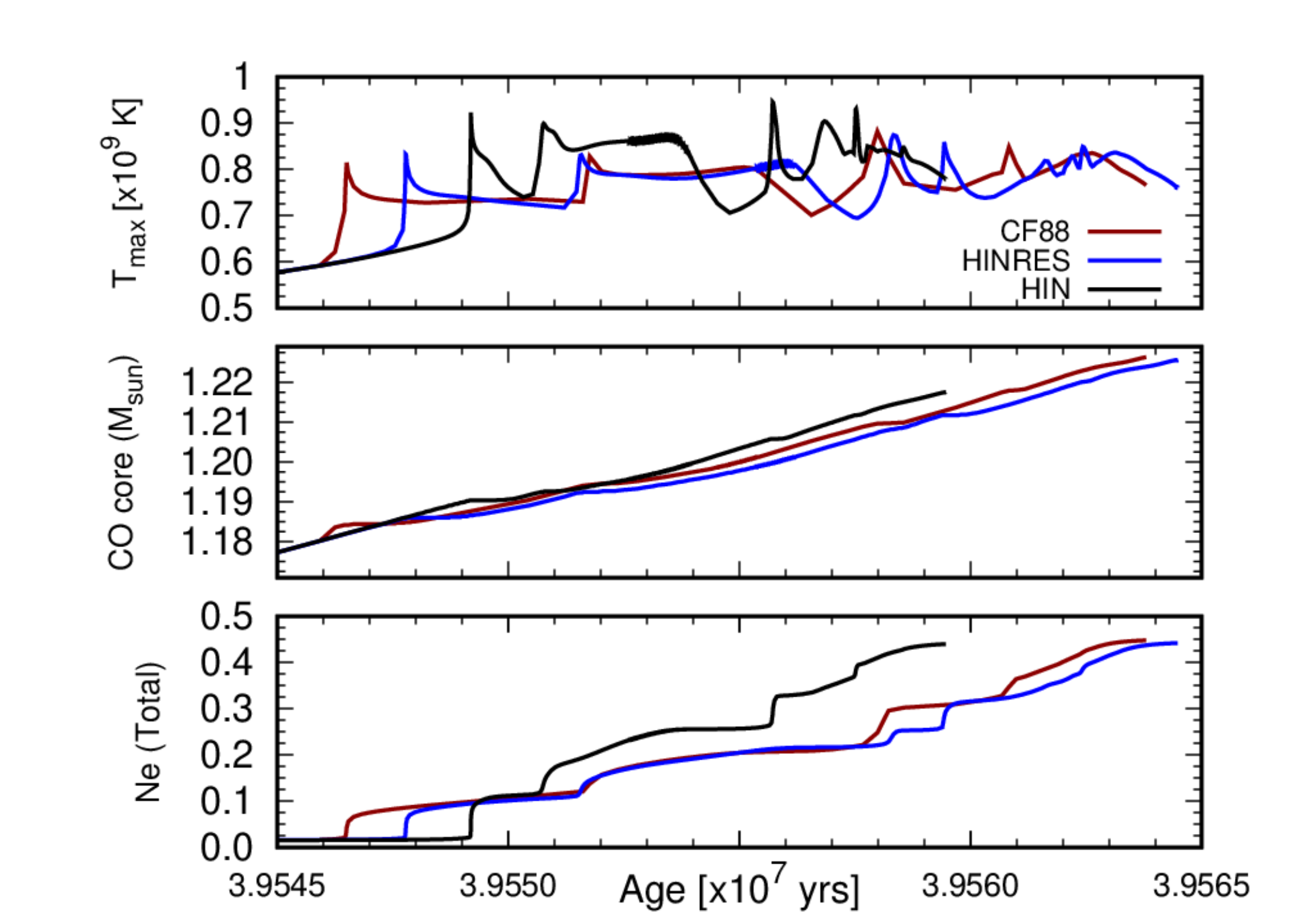}
   \caption{Run of the maximum temperature of the flame (upper panel), the size of the CO core (middle panel), and the total $^{20}$Ne content left (lower panel) for all the reaction rates adopted for a $M_{\rm ZAMS}=8 \, M_{\odot}$ model during the lapse of the carbon burning phase.
   }\label{fig:evol}
\end{figure}

     \begin{figure}
  \includegraphics[width=1\columnwidth]{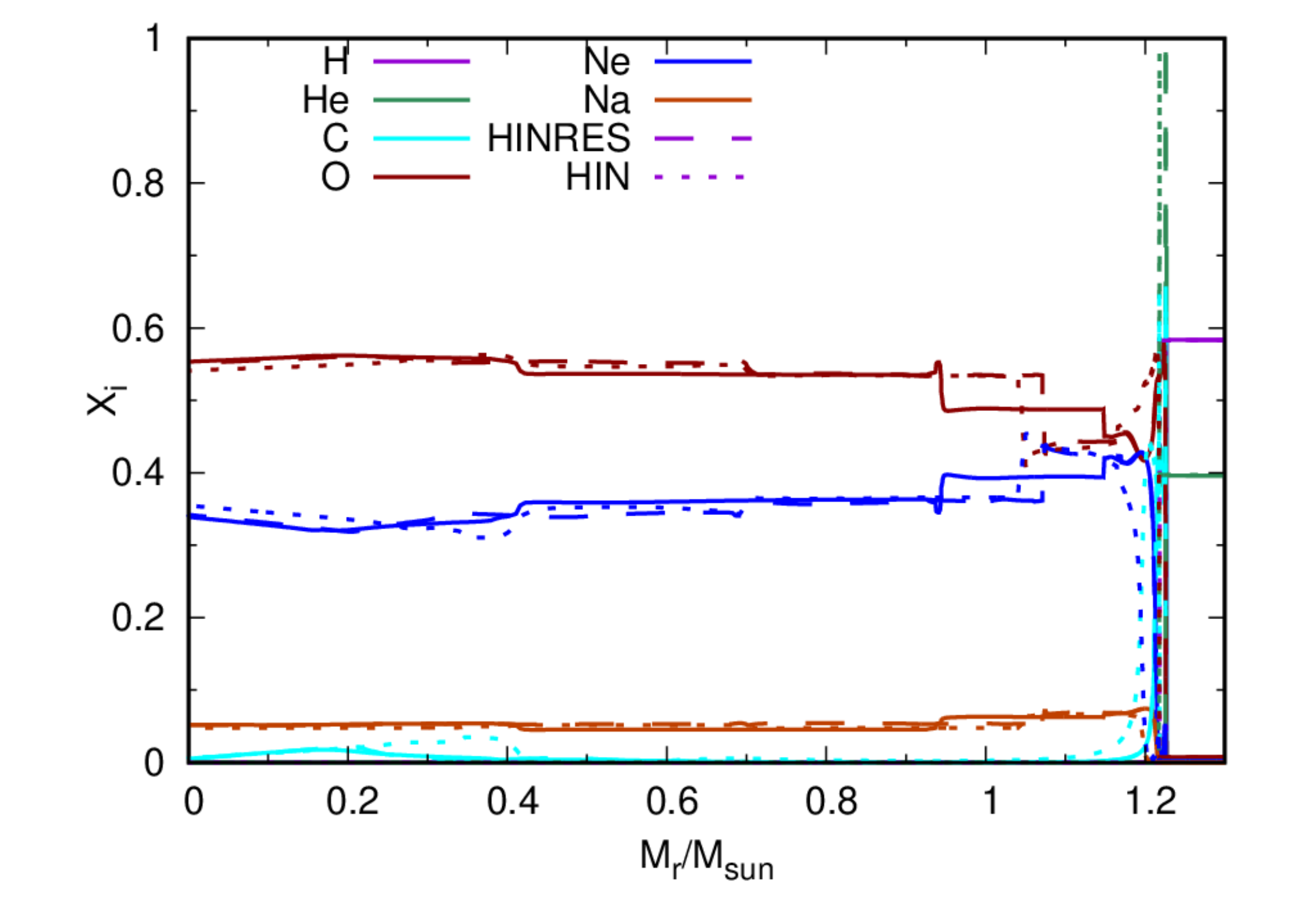}
   \caption{Chemical profiles for the most abundant elements at the end of the carbon burning phase for a $M_{\rm ZAMS}=8 \, M_{\odot}$ model for the three nuclear reaction rates adopted, namely CF88 (solid lines), HINRES (dashed lines), and HIN (dotted lines). The branching ratio was set to 65\% and 35\% for the $\alpha$ and $p$ channels, respectively. In this plot, $X_i$ is the element abundance per mass, whereas $M_r$ is the Lagrangian mass coordinate.}
  \label{fig:prfile-8}
\end{figure}

\section{Stellar models}
\label{sect:models}
The SAGB models employed in this work were computed with 
 the stellar evolution code Modules for Experiments in Stellar Astrophysics ({\tt MESA}) version r21.12.1 \citep{Paxton2011, Paxton2013, Paxton2015, Paxton2018, Paxton2019}.  Most of the adopted input physics corresponds to the default options that are described in detail in those papers, and will thus not be repeated here. The nuclear network adopted  (sagb\_NeNa\_MgAl.net) accounts for 28 isotopes n, $^{1,2}$H, $^{3,4}$He, $^{7}$Li, $^7$Be, $^8$B, $^{12,13}$C, $^{13-15}$N, $^{16-18}$O, $^{19}$F, $^{20-22}$Ne, $^{21-23}$Na, $^{24-26}$Mg, $^{25-27}$Al, including all the dominant nuclear reactions. Most of these reactions are taken from CF88 and \citet{1999NuPhA.656....3A}, with some exceptions.
 Convective boundary mixing (CBM) was only adopted during core H and He burning, with the suggestion of \citet{1996A&A...313..497F} of an exponentially decaying velocity field.  Here, the diffusion coefficient is adopted according to \citet{1997A&A...324L..81H}, with the free parameter $f=0.017$.  The main impact of CBM in these stages is to decrease the initial mass required for a progenitor star to reach C-ignition by about $2\, M_\odot$.
 CBM at the C-burning convective zones and at the bottom of the convective envelope was not included.
 
Our evolutionary sequences were computed from the zero-age main sequence (ZAMS) through central hydrogen and helium burning, up to the end of the carbon burning stage, before the thermally unstable phase, for models with initial masses $6.85\leq M_{\rm ZAMS}/M_{\odot}\leq 8.50$ and metallicity $Z=0.02$, assuming a solar-scaled composition. Each model was computed with a spatial resolution greater than 6000 spatial points from center to surface, while the temporal resolution achieved is of the order of 14 yrs, similar to those adopted in \cite{2015ApJ...807..184F}.
The evolutionary behavior of SAGB progenitors before carbon
ignition is very similar to that of intermediate-mass stars that end
up as CO WDs, well documented in previous works \citep{1994ApJ...434..306G, 2006A&A...448..717S, 2010A&A...512A..10S}. 
The development of carbon burning, under the hypothesis of a strict Schwarzschild criterion \citep{1906WisGo.195...41S} for the delimitation of the convective region, is characterized by two different stages \citep{1994ApJ...434..306G, 2006A&A...448..717S}. The first corresponds to the ignition of C at the point of maximum temperature inside the partially degenerate CO core, inducing a thermal runaway called the carbon flash. The sudden energy injection by the C-flash leads to a convective zone, which extends outward from the point of maximum temperature. The second stage corresponds to the development of a flame that propagates to the center and transforms the CO core into an ONe core \citep{1994ApJ...434..306G,  2006A&A...448..717S}. During the C-burning phase, we took into account three different reaction rates for the $^{12}$C+$^{12}$C reaction, namely the aforementioned HIN and HINRES ones from \cite{2022A&A...660A..47M}  and the value from CF88, which has been widely used in previous evolutionary computations. Mass loss during AGB was considered adopting the Bloecker's wind prescription with a scaling factor of 0.1 \citep{1995A&A...297..727B}. 

 The adoption of the Schwarzschild Criterion for convective instability in our pre-WD models implies that Rayleigh-Taylor unstable regions are not identified as such in our models. In our models, the CO-core during the early AGB develops a off-centered peak in the oxygen profile \citep{1997ApJ...486..413S}. Such a profile would be unstable to Rayleigh-Taylor instabilities \citep{1977ApJS...35..239S}. A proper assessment of mixing processes driven by chemical gradients during the early- and TP-AGB phases would require the computation of these processes during the formation of the oxygen peak. For simplicity, in our computations, the chemical profile of the CO core is homogenized by an ad-hoc algorithm just before the WD cooling phase.

\begin{figure}
    \includegraphics[width=1\columnwidth]{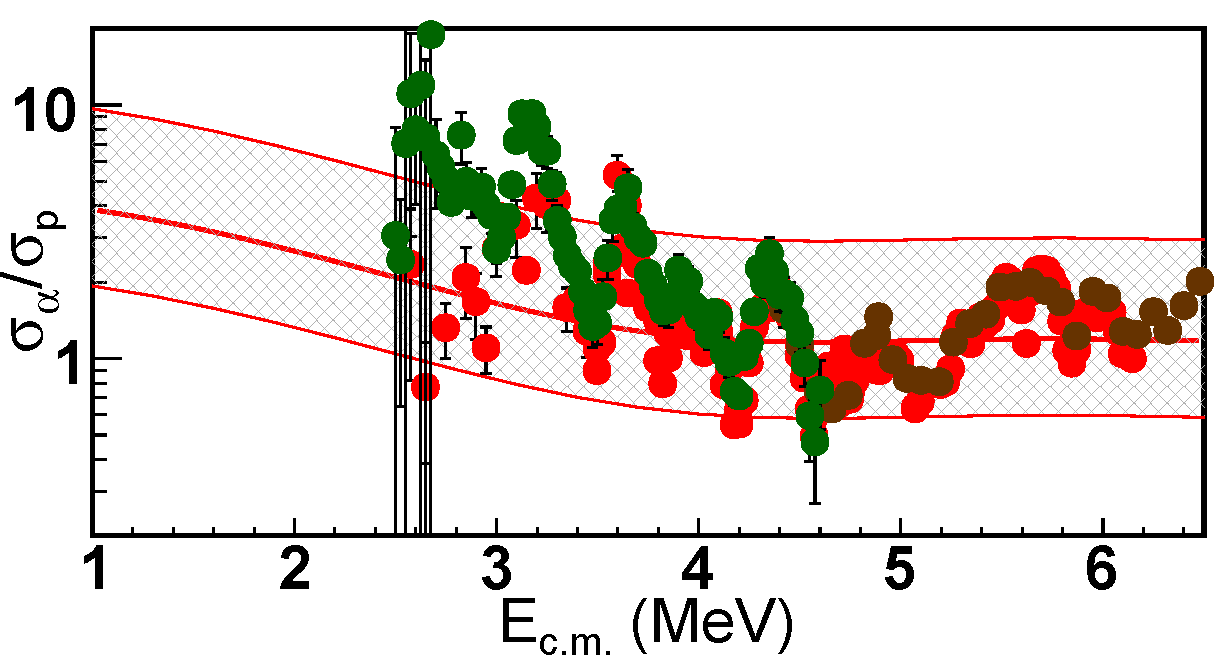}

   \caption{The $S^*$ ratio of the alpha and proton channels.  The measurements of  \cite{1980ZPhyA.298...65K}, \cite{2006PhRvC..73f4601A}, and \cite{2007PhRvL..98l2501S} are shown as filled circles in the colors of dark green, red, and brown, respectively. The best fit, upper and lower limits are shown as red lines. These limits are set at factors of 2.5 and 0.5 of the best fit, respectively}
  \label{fig:alpha-p}
\end{figure}

\begin{figure}
  \includegraphics[width=1\columnwidth]{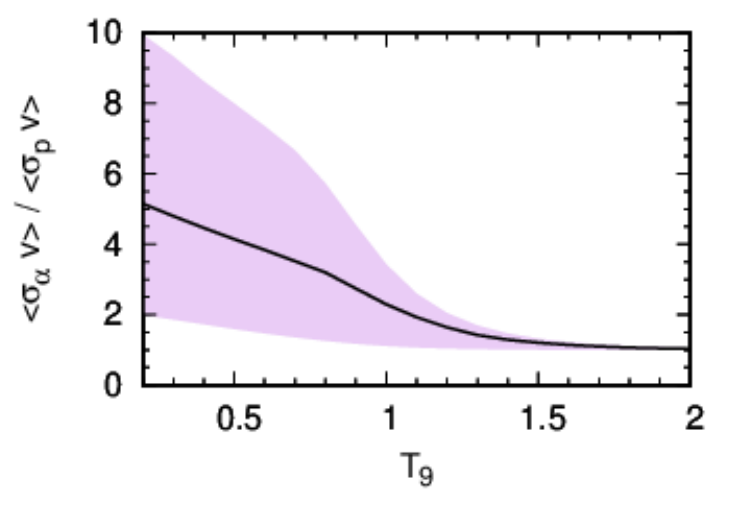}
   \caption{Ratio of the $^{12}$C$(^{12}$C,$\alpha)^{20}$Ne to the $^{12}$C$(^{12}$C,p$)^{23}$Na channel as a function of temperature. The color band indicates the region between the estimated upper and lower limits.}
  \label{fig:branching}
\end{figure}

 \section{Impact on the stellar structure}
 \label{sect:impact-ste}

\subsection{From uncertainties in the total rate}
\label{unc-rate}

 In this section, we computed the evolutionary sequences adopting different reaction rates with a fixed branching ratio of 65\% for the $\alpha$ channel.
 In Fig. \ref{fig:evol}, we show the impact of the differences in the reaction rates shown in Fig. \ref{fig:rate} on the properties of carbon burning. Specifically, we show the evolution of the maximum temperature of the flame ($T_{\rm max}$, upper panel), the size of the CO core (middle panel), and the total $^{20}$Ne produced (lower panel) as a function of the age, for an $8 \, M_{\odot}$ progenitor model. 
 As it is known, in the case of burning shells on top of electron-degenerate cores, the temperatures of the burning shell and the core increase with an increment of the size of the core \citep{2022ApJ...941..149M}. Sequences computed with a higher $^{12}$C+$^{12}$C reaction rate require slightly lower core temperatures to ignite carbon and undergo the carbon flash, which thus takes place at lower core masses and temperatures. Conversely, the sequences computed with the less efficient reaction rates are characterized by a late onset of the C-flash, larger cores at the moment of the C-flash, and higher C-burning temperatures, as also seen in the study of more massive stars \citep{2012MNRAS.420.3047B, 2013ApJ...762...31P}. 
Once carbon starts to burn quiescently, more massive cores are forced, by hydrostatic equilibrium, to burn carbon faster. Consequently, those sequences with lower $^{12}$C+$^{12}$C reaction rates that ignited carbon with slightly more massive cores display shorter carbon-burning lifetimes.
  Differences in the size of the cores at the moment of the C-flash, in their $T_{\rm max}$, and in the lifetime of the C-burning phase amount to 0.5, 13, and 40\%, respectively.

Despite the differences in the burning temperature and the efficiency of the reaction rate, the total $^{20}$Ne content for each sequence is practically identical, as seen in the lower panel of Fig. \ref{fig:evol}. This is also observed with the distribution of the most important elements (see Fig. \ref{fig:prfile-8}). The most noticeable differences in the chemical structure arise in the outermost part of the core, as shaped at the final stages of C-burning. However, as a general trend, differences in the resulting distribution of elements are small.

 In the next sections, we will study the impact of such differences on the age, crystallization, and pulsation properties of ultramassive DAV stars, which are H-rich WDs that present $g$-mode pulsations \citep{2019A&ARv..27....7C,2015pust.book.....C}.
 Before, we will analyze the uncertainties coming from the branching ratios of the $^{12}$C$+^{12}$C nuclear reaction.
 \label{sect:impact}

  \subsection{From uncertainties in the branching ratios}
 \label{sect:impact-br} 
As mentioned in Sect.~\ref{sec:int},  the $^{12}$C$(^{12}$C,$\alpha)^{20}$Ne and $^{12}$C$(^{12}$C$,p)^{23}$Na reactions dominate carbon fusion inside stars, leaving $^{20}$Ne and $^{23}$Na as the main end products. 
The probability of each exit channel becomes very uncertain for the typical temperatures at which C-burning takes place in stars \citep{2013ApJ...762...31P,2022EPJWC.26001002T}. 


From the measurements made by \cite{1980ZPhyA.298...65K}, \cite{2006PhRvC..73f4601A}, and \cite{2007PhRvL..98l2501S},  we assessed the estimation of the ratio of the $\alpha$- to the $p$-channel, $S^*_{\alpha}/S^*_{p}$, after correcting for the branching ratio of the missing channels with the procedure described in \cite{2020ChPhC..44k5001L}. The theoretical prediction is calculated using TALYS \citep{2005AIPC..769.1154K}. The TALYS prediction is scaled by a factor 1.33 to match with the average of the experimental data. The results are shown in Fig. \ref{fig:alpha-p}.
We recommend upper and lower limits to be set by scaling the best fit with factors of 2.5 and 0.5, respectively. These limits covers all the measurements above $E_{\rm c.m.}=$ 3.3 MeV. At lower energies, the measurement of \cite{1980ZPhyA.298...65K}  agrees with the limits while the measurement of \cite{2007PhRvL..98l2501S} becomes significantly higher than that of \cite{1980ZPhyA.298...65K} and exceeds the upper limit by a factor of 2.7 around $E_{\rm c.m.}>$ 3.16 MeV. Further experimental and theoretical investigations are needed to resolve the discrepancy.  

The reaction rates of the $\alpha$- and $p$-channels are calculated based on the extrapolated and experimental $S^*$ factor or cross-sections.  At $E_{c.m.}>$ 2.7 MeV, $S^*_{\alpha}$ and $S^*_{p}$ obtained from the experimental measurements are used after correcting the missing channels in their measurement.  $S^*_n$ is estimated with the experimental data and theoretical extrapolation from \cite{2015PhRvL.114y1102B} and \cite{2014PhDT........67B}.  At lower energies, the $S^*$ factor of $^{12}$C+$^{12}$C is generated with a uniform distribution bound by the upper and lower limits recommended in \cite{2020ChPhC..44k5001L}. $S^*_{\alpha}$ and $S^*_p$ are generated with the branching ratios of the $\alpha$- to $p$-channels with a uniform distribution bound by the upper and lower limits shown in Fig. \ref{fig:alpha-p}. The reaction rate ratio of the Ne and Na channels are shown in Fig. \ref{fig:branching}. Our results show a decrease with temperature by about 40\% from $T_9=0.6$ to $T_9=0.9$, and are significantly higher than the value ${\alpha}/{p}$=1.27 recommended by CF88, at the temperatures of interest.

To quantify the impact of these uncertainties on the chemical composition of stars at the end of the SAGB phase, we performed calculations considering a wide range of branching ratios.
To this end, we computed evolutionary sequences for stellar models with masses $6.80\leq M_{\rm ZAMS}/M_{\odot}\leq 8.50$ considering the CF88 and HIN reaction rates and, from the results shown in Fig. \ref{fig:branching} that indicates $\alpha/p>1$ in the range of temperatures of interest, branching ratios of $[\alpha/p]$ of [90/10], [80/20], [65/35], [60/40], [55/45], and [50/50].
 Higher branching ratios for the $\alpha$ channel translate into smaller initial masses needed for the occurrence of the C-flash (by $<1.5\%$). This happens because the total energy released per burned $^{12}$C nuclei by the $\alpha$ channel is higher. Consequently, a lower burning rate is required to reach the carbon-burning luminosity at which the runaway occurs. 
 This means that the core ignites carbon at a slightly lower core 
 temperature, and consequently a smaller core mass. These differences are minor, and the onset of the C-burning flash occurs almost at the same age, and differences in the sizes of the cores are at most of 0.1\% Conversely, the lifetime of the C-burning phase increases by up to 20\% when the $\alpha$ channel is more efficient due to the larger energy released per burnt nuclei. 
 
\begin{figure}
    \includegraphics[width=1\columnwidth]{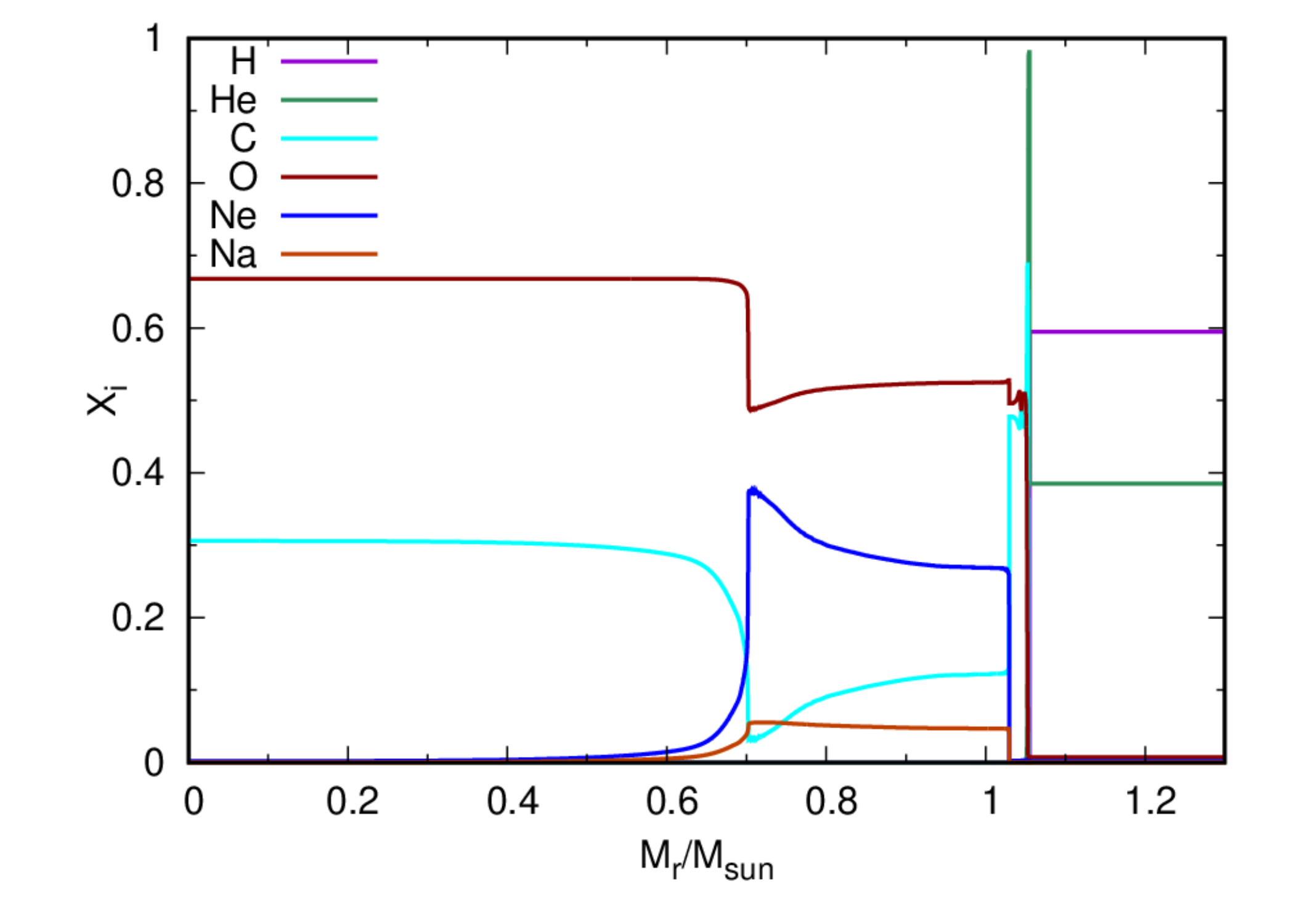}

   \caption{Chemical structure for a $6.85 \, M_{\odot}$ model computed with the CF88 reaction rate, at the end of the C-burning phase. The flame is quenched before reaching the star's center, leaving a core composed of CO surrounded by an ONe mantle.}
  \label{fig:profile-685-65-35}
\end{figure}

 For some of our sequences, the flame is quenched before reaching the star's center. This means that $^{12}$C does not burn completely inside these objects and therefore the chemical structure left at the end of this stage is that of a core composed of CO surrounded by an ONe mantle. This depends on the reaction rate adopted and its branching ratio. Particularly, if the CF88 (HIN) reaction rate is adopted with the [65/35] branching ratio, this happens for models with initial masses $6.85\leq M_{\rm ZAMS}/M_{\odot} < 7.10$ ($7.25< M_{\rm ZAMS}/M_{\odot} < 7.40$). The resulting object will be a  CO-core or hybrid\footnote{In the following we will refer as hybrid those stars with a core composed by $^{12}$C, $^{16}$O and a significative amount of $^{20}$Ne.} WD, depending on the total amount of $^{12}$C burned.
 In Fig. \ref{fig:profile-685-65-35}, we show the chemical structure for a $M_{\rm ZAMS}=6.85 \, M_{\odot}$ model.
 An unburnt CO core is surrounded by an ONe mantle similar to those found when overshooting is adopted during the C-burning phase \citep{2013ApJ...772...37D,2022A&A...659A.150D}.  However, 3D hydrodynamic simulations revealed that this quenching of the carbon flame might be an artificial consequence of overshooting prescriptions adopted in 1D models
\citep[see][]{2016ApJ...832...71L}. Conversely, our computations do not include any extra-mixing process during the C-burning stage, and consequently, the quenching of the C-flame is of a different nature. Depending on the size of the ONe mantle, i.e. the total amount of $^{20}$Ne produced, the chemical structure at the WD phase, post Rayleigh-Taylor homogenization, could be a WD with a core composed mostly of CO (see next sections). 
It should be noted that the initial mass ranges for the different regimes discussed in this paper are dependent on the assumptions on convective boundary mixing. More extended convective boundary mixing in the hydrogen burning core on the main sequence would imply lower initial masses for C-ignition \citep{2020MNRAS.493.4748W}. Convective boundary mixing at the carbon-burning flame is also expected to strongly affect the development of the carbon flash \citep{2013ApJ...772...37D,2022A&A...659A.150D}.

\begin{figure}
  \includegraphics[width=1\columnwidth]{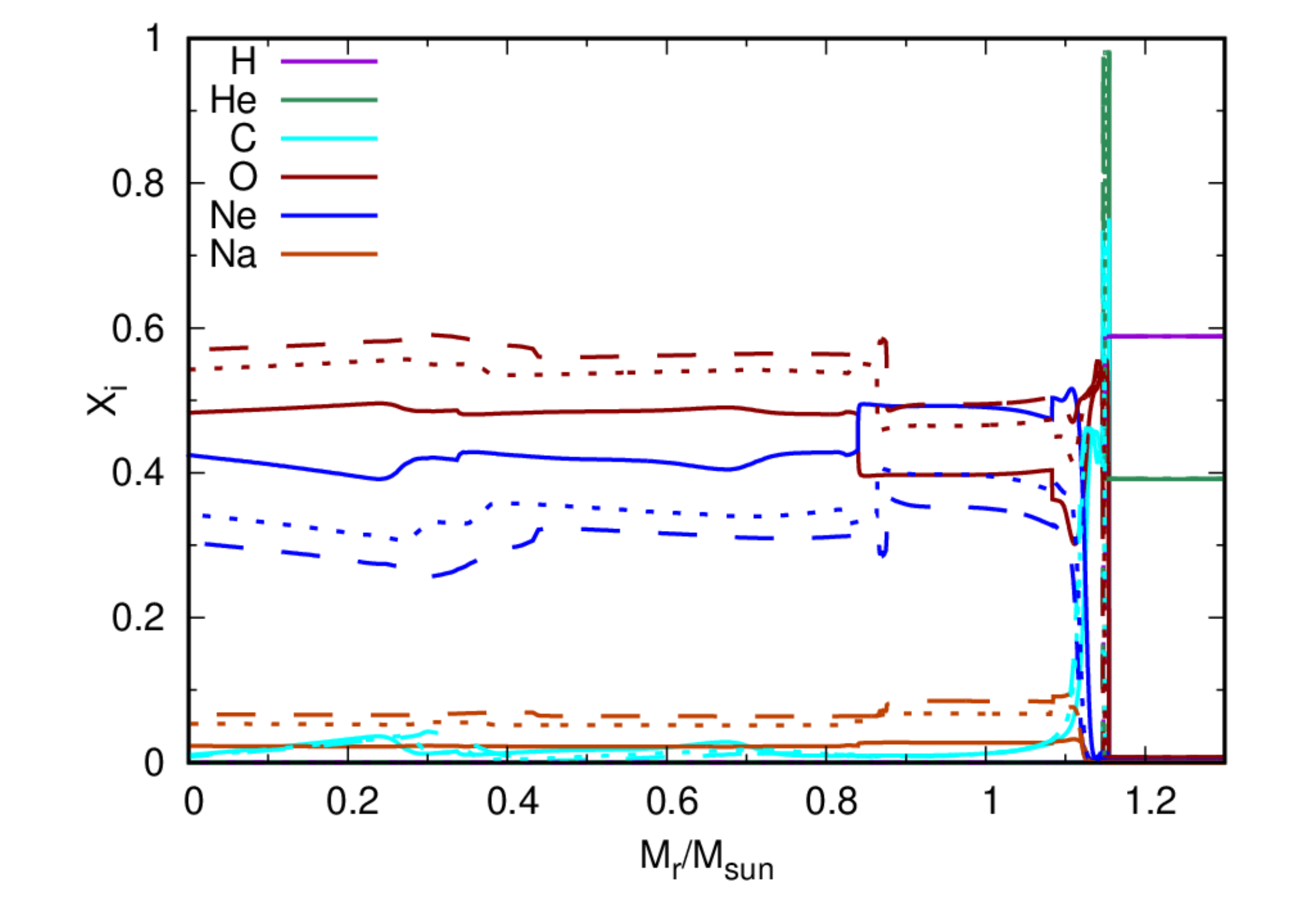}
   \caption{Chemical structure for a $7.50 \, M_{\odot}$ model at the end of the C-burning phase, for models adopting the CF88 reaction rate with [55/45],  [65/35] and [90/10] branching ratios (dashed, dotted, and solid lines, respectively).}
  \label{fig:profile-750-branches}
\end{figure}

 In Fig. 
 \ref{fig:profile-750-branches}, we show the chemical profiles for the $7.50 \, M_{\odot}$ model at the end of the C-burning phase, assuming the branching ratios adopted in the literature, namely [55/45] (CF88) and [65/35] \citep{2013ApJ...762...31P}, together with the extreme case [90/10]. 
 Depending on which of these branching ratios is selected, differences in the central abundances of O and Ne can reach from 3\% to 17\%. The larger differences arise when comparing the calculations adopting the [55/45] and [90/10] branching ratios.
 For the [90/10] branching ratio, the $^{20}$Ne production exceeds that of $^{16}$O in the outermost part of the core ($M_r/M_{\odot}>0.8$). This is because, as it moves inwards, the temperature the C-burning decreases. These differences in the core abundances of $^{16}$O and $^{20}$Ne will be critical for the crystallization process during the WD cooling phase.

\section{Consequences for the properties of WDs}
\label{sect:consequence}

This section is devoted to analyzing the effect of the different assumptions studied previously on the chemical structure and composition of ultramassive WDs, their cooling timescales, crystallization, and pulsation properties.

Each WD evolutionary sequence was computed from the point of maximum temperature of the cooling track at high luminosities, down to the development of the Debye cooling at low surface luminosities.
 Initial WD structures were constructed by mapping the detailed chemical structure of the H-free core at the end of the central C-burning stage into an already existing WD thermal structure. Our DA WDs models have a H content of $M_{\rm H}\sim 1.5\times 10^{-6}M_{\rm WD}$.
The evolution and structure of the WD models presented in this section were computed with the {\tt LPCODE} evolutionary code \citep[for details, see][]{2003A&A...404..593A,2005A&A...435..631A,2015A&A...576A...9A,2021A&A...646A..30A,2016A&A...588A..25M}. During crystallization, we took into account the release of latent heat and changes in the core chemical composition resulting from phase separation upon crystallization, using phase diagrams suitable for C/O or O/Ne plasmas 
\citep{2019A&A...625A..87C,2022MNRAS.511.5198C}. For the sake of clarity, we selected a fiducial model of initial mass $7.50 \, M_{\odot}$, which corresponds to a $M_{\rm WD}\sim 1.15 \, M_{\odot}$. 

As discussed in the previous sections, the adoption of extreme reaction rates (CF88 and HIN) during the C-burning phase leads to small differences in the chemical structure of the ultramassive WD progenitors at the end of the C-burning phase. The next evolutionary stage is the thermally unstable phase and consists of the build-up of the most external part of the core.
After the WD starts to cool and the Rayleigh-Taylor re-homogenization has taken place, the chemical structures of the WDs computed with different nuclear reaction rates are almost indistinguishable.
As the WD keeps cooling, crystallization gets underway and, by the time the star reaches the ZZ Ceti (DAV) instability strip (at a temperature of $\sim$ 12\,000~K), the cooling times and the size of the crystallized core are of 1743 (1721) $\times 10^6$~yr and 92.9 (92.2) \%, respectively, if the CF88 (HIN) rate is adopted, meaning small differences of  0.5 and 1.2\%.

\begin{figure}
  \includegraphics[width=1\columnwidth]{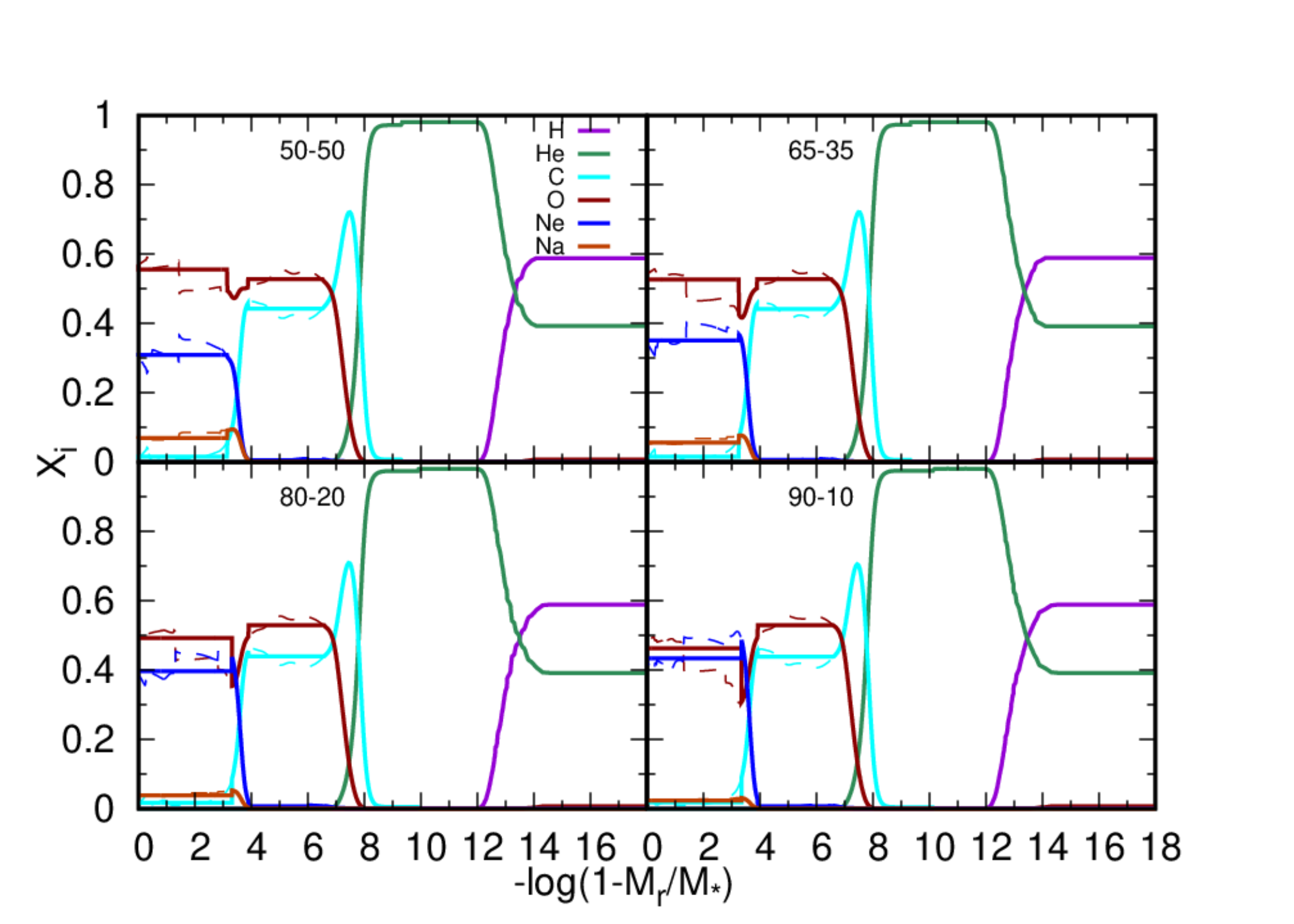}
  \includegraphics[width=1\columnwidth]{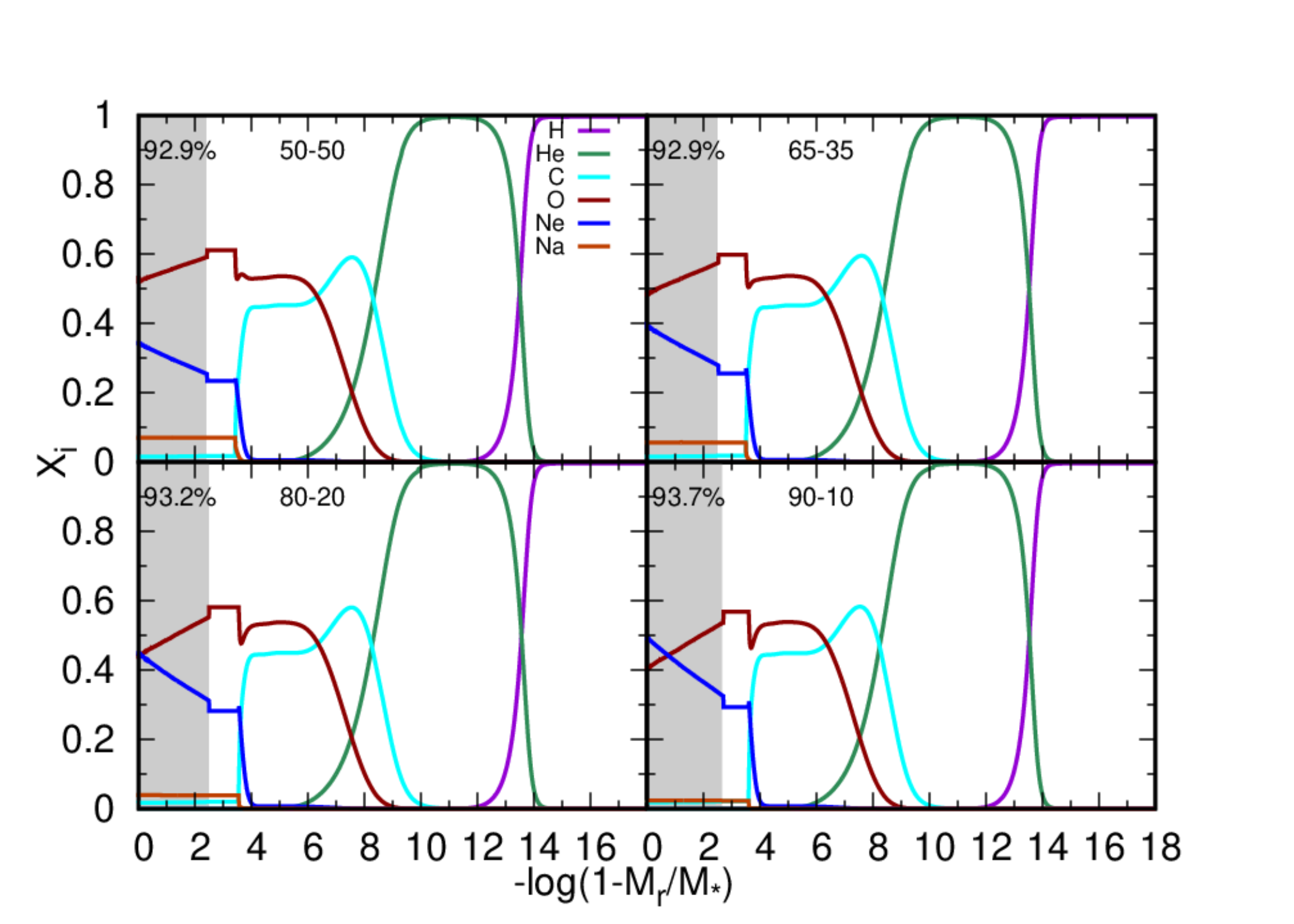}

  \caption{Distribution of the most important chemical elements for the $M_{\rm ZAMS}=7.50 \, M_{\odot}$ models computed with the CF88 rate and different branching ratios (from upper left, in clockwise order, 50-50, 65-35, 90-10, and 80-20) at the beginning of the WD's cooling path ($\sim 10^5$~K), pre- and post-Rayleigh-Taylor rehomogenization (upper panel, dashed and filled lines respectively) and at the ZZ Ceti stage ($\sim 12\,000$~K, lower panel). The grey region reflects the crystallized part of the star, with a crystallization level above 90\%.}
  \label{fig:profile-pre-postRT}
\end{figure}

Regarding the adoption of different branching ratios, the differences found in the distribution of the most important chemical elements are more noticeable. In the top panel of Fig. \ref{fig:profile-pre-postRT}, we show the chemical structure in terms of the outer mass fraction $\rm log(1-M_r/M_*)$\footnote{ $\rm M_r$ and $\rm M_*$ stands for the mass coordinate and total mass of the star, respectively.} pre- and post-Rayleigh-Taylor re-homogenization, at high temperatures, for the fiducial model that accounts for the following branching ratios: [50/50], [65/35], [80/20], and [90/10]. The central  $^{23}$Na content remains always below $\sim$ 8\% and, as higher production rates are adopted, the $^{20}$Ne central content increases from 30\% up to $\sim$ 50\% for the [90/10] branching ratio. In the latter case, the $^{20}$Ne content produced at the outer part of the core surpasses the $^{16}$O content before rehomogenization. However, the final chemical structure of the WD is that of a typical ONe core WD, but composed of almost equal parts of $^{16}$O and $^{20}$Ne.
The C/O mantle on top of the O/Ne core is identical for each case. 

 In the lower panel of Fig. \ref{fig:profile-pre-postRT}, we show the same chemical profiles but at the ZZ Ceti stage ($T \sim 12\,000$~K, when the crystallization of the core has reached more than 90\%, indicated as a shaded region in the plot). As every model has the same mass, the differences in the percentage of the crystallized core, which amounts to up to 0.8\%, come strictly from the differences in their composition. The combined effect of the different degrees of crystallization and inner composition leads to differences in the WD cooling times of at most 1.3\%.

\begin{figure}
  \includegraphics[width=1\columnwidth]{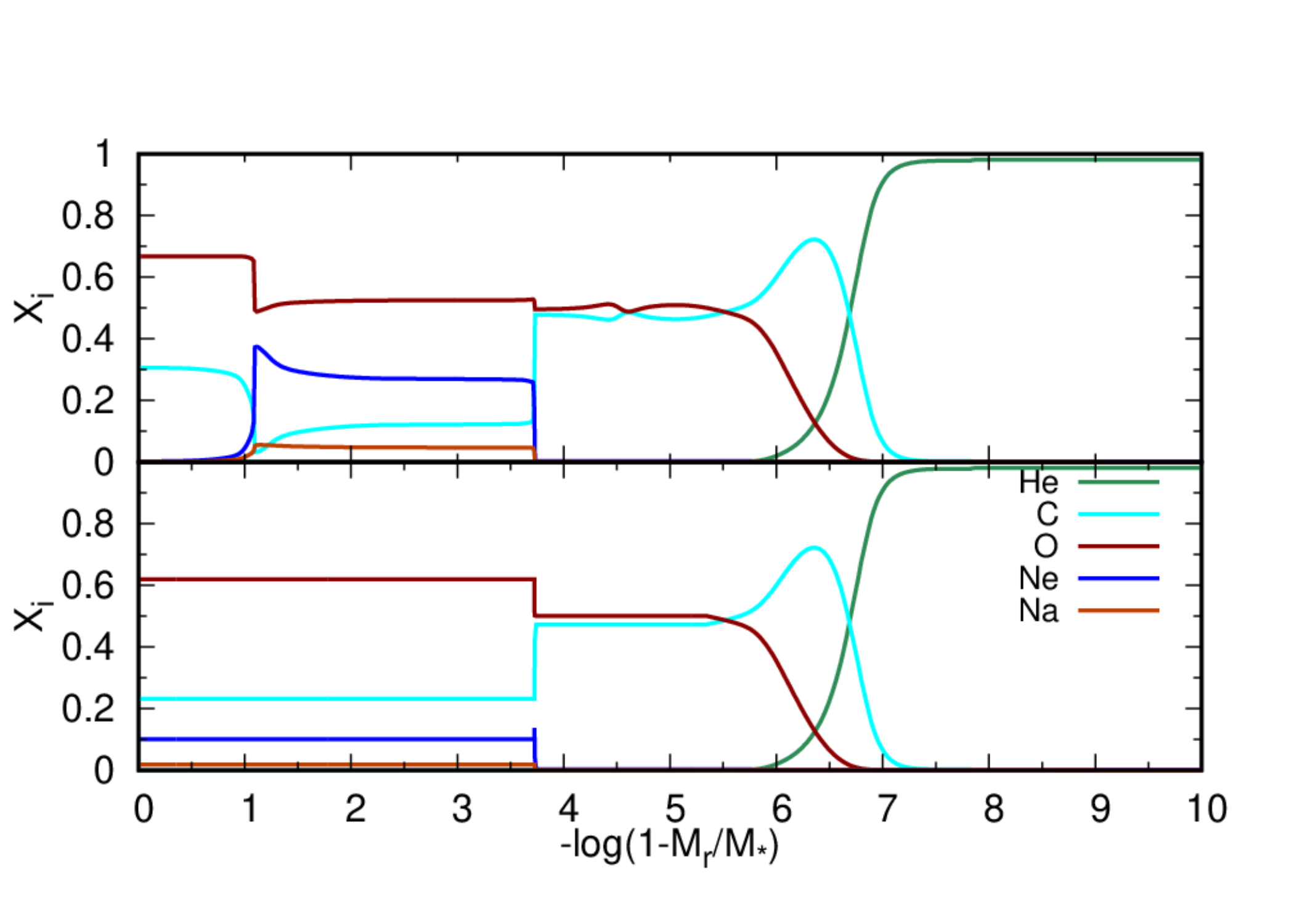}
   \caption{Distribution of the most abundant elements for a hybrid CONe-core model at $10^5$~K, pre- and post-Rayleigh Taylor rehomogenization (upper and lower panel, respectively).}
  \label{fig:profile-685}
\end{figure}

\begin{figure}
  \includegraphics[width=1\columnwidth]{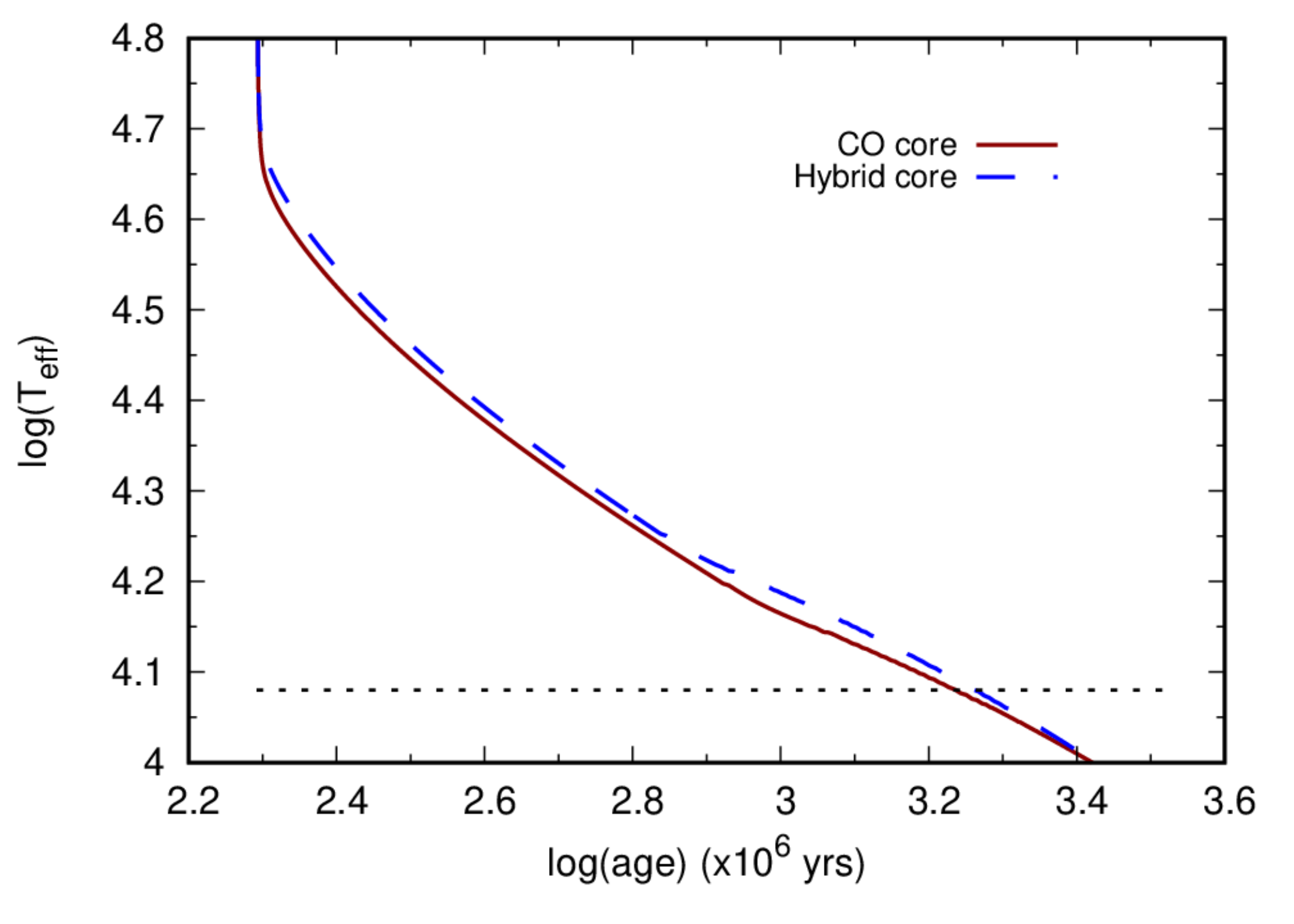}
   \caption{Evolution of the effective temperature $T_{\rm eff}$ as a function of the age, for both the pure $M_{\rm ZAMS}=6.85M_{\odot}$ CO core and hybrid models. The black horizontal dotted line indicates a $T_{\rm eff}=12\,000$~K.}
  \label{fig:teff-evol}
\end{figure}

As noted in Section \ref{sect:impact-br}, under some of the conditions discussed in the present paper, the carbon flame is quenched prematurely, forming a hybrid core WD.
   In Fig. \ref{fig:profile-685}, we show the chemical structure for the particular case of a hybrid core, with $M_{\rm WD}/M_{\odot}=1.05$ ($M_{\rm ZAMS}/M_{\odot}=6.85$) pre- and post-Rayleigh-Taylor re-homogenization (upper and lower panel, respectively).  The total $^{20}$Ne content produced during the short-lived C-burning phase is then redistributed (as are the other species) throughout the core. Consequently, after re-homogenization,  the WD model's structure resembles that of a pure CO-core WD, though with a non-negligible $^{20}$Ne content ($\sim$10\%) in the core. These new kinds of objects predicted by our study, which can have masses up to $M_{\rm WD}/M_{\odot}\sim 1.11$, differ substantially from objects with pure CO cores. The presence of $^{20}$Ne modifies the structure of the WD (particularly its central density) so that differences in the evolution are significant. This can be seen in Fig. \ref{fig:teff-evol}, where we show the evolution of the effective temperature for a pure CO and the hybrid model.\footnote{The pure CO-core model was computed from the evolution of the hybrid CONe-core model progenitor but quenching C-ignition.} Both models start at the same evolutionary point in the cooling track but differ in their central compositions, which are 30\%-67\% ($^{12}$C-$^{16}$O) for the pure CO-core model and 23\%-61\%-10\% ($^{12}$C-$^{16}$O-$^{20}$Ne) for the hybrid model.
   As the models cool down, the hybrid one starts crystallizing earlier than the CO-core model. The crystallization onset occurs at 751$\times 10^6$ yr for the hybrid composition, while for the pure CO-core model, it starts at 870$\times 10^6$ yr. At 12\,000 K (grey line), the age differences are of $\sim6\%$. 

\begin{figure}
  \includegraphics[width=1\columnwidth]{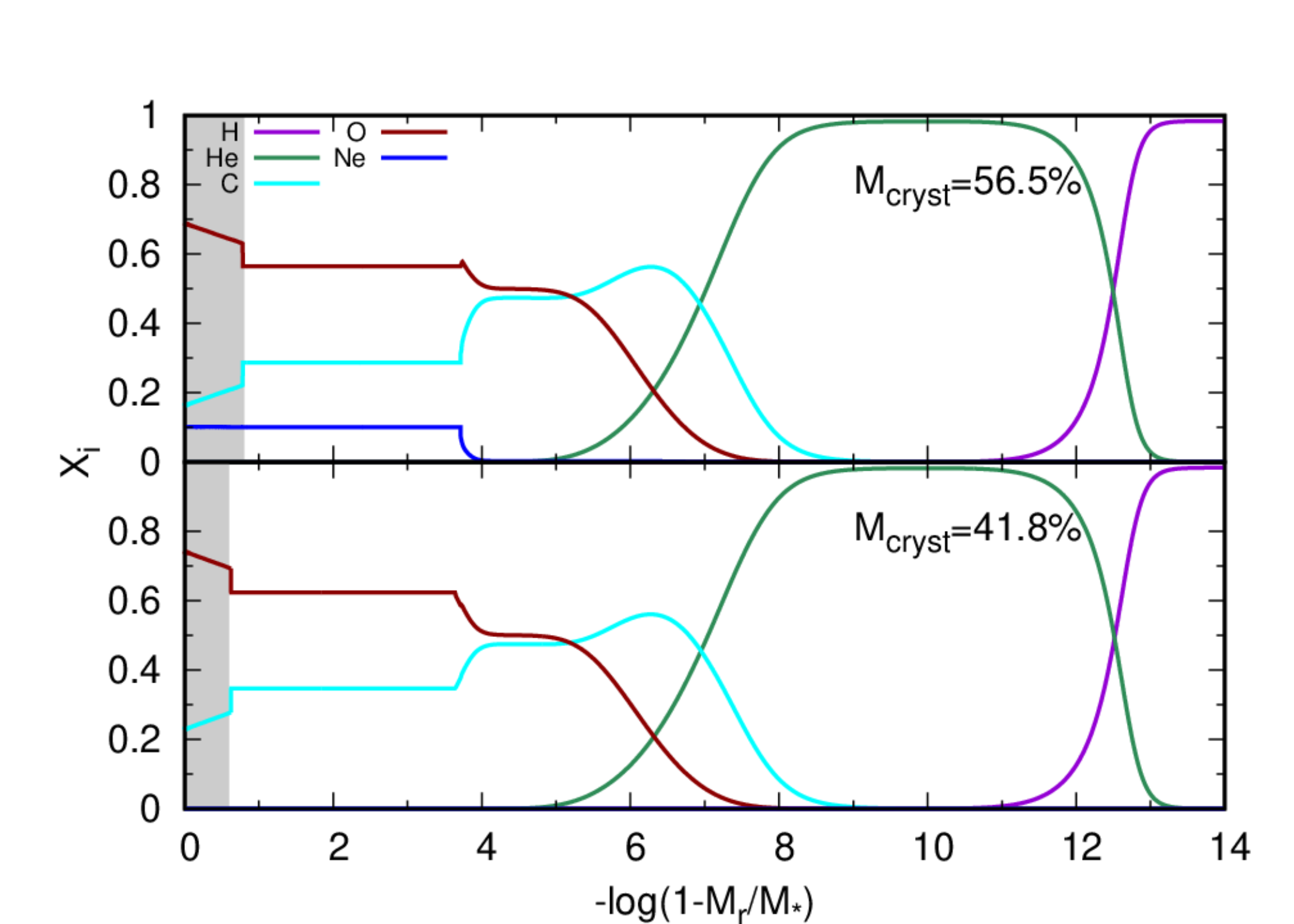}
   \caption{Chemical profile of the $M_{\rm ZAMS}=6.85 \, M_{\odot}$ hybrid and pure CO-core models (upper and lower panel, respectively) at 12\,000~K. The shaded region depicts the crystallized portion of the core, which reach crystallization levels of 56.5 and 41.8\%, respectively.}
  \label{fig:profile-685-12000}
\end{figure}

The comparison of their chemical structures at this point can be seen in Fig. \ref{fig:profile-685-12000}, where we show the distribution of the most important chemical species for both models. Both the absence of $^{20}$Ne in the pure CO-core model and the crystallized size of the core are the most prominent differences at this point, while the $^4$He-buffer region and pure $^1$H-envelope remain the same.


   \begin{figure}
  \includegraphics[width=1\columnwidth]{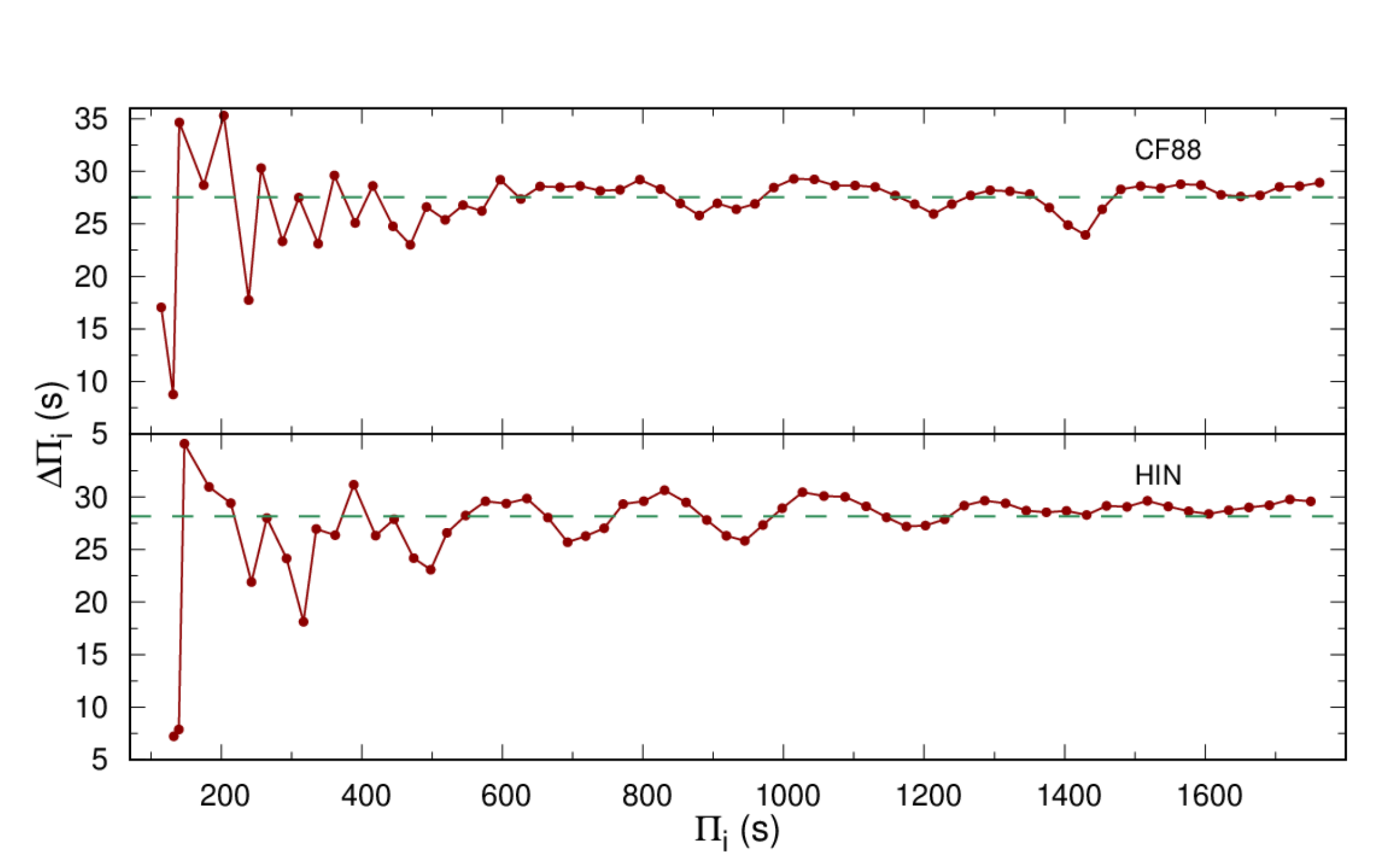}
   \caption{Comparison of the period spacing $\Delta \Pi_i$ as a function of the periods of the modes $\Pi_i$ for the CF88 and HIN models (upper and lower panel, respectively). The green dashed line shows the value of the asymptotic period spacing.}
  \label{fig:dp-rate}
\end{figure}

\subsection{Asteroseismology}

The period spectrum and mode-trapping properties of $g$-modes of DAV WDs depend sensitively on the precise shape of the Brunt-V\"ais\"al\"a frequency across the interior of the star, and, in particular, the location and shape of the bumps produced by the chemical composition interfaces \citep{2019A&A...621A.100D, 2010A&ARv..18..471A}. Thus, any change in the chemical profiles
translate into changes in the WD's expected pulsation properties. 

The computation of the pulsation properties for all our ultra-massive pulsating DA WD models was done using the {\tt LP-PUL} pulsation code described in \citet{2006A&A...454..863C}, previously employed in the study of the properties of ultra-massive WD models \citep{2019A&A...621A.100D, 2022A&A...659A.150D} and used to perform asteroseismic studies of ultra-massive ZZ Ceti stars \citep{2019A&A...632A.119C,2023MNRAS.522.2181K}. Element diffusion was included for all models from the beginning of the WD cooling track. 
 The “hard sphere”  boundary conditions were adopted when accounting for the effects of crystallization on the pulsation properties of the $g$-modes. These conditions assume that the amplitude of the eigenfunctions of $g$-modes is drastically reduced below the solid and liquid interface, as compared with the amplitude in the fluid region \citep{1999ApJ...526..976M}.

    \begin{figure}
  \includegraphics[width=1\columnwidth]{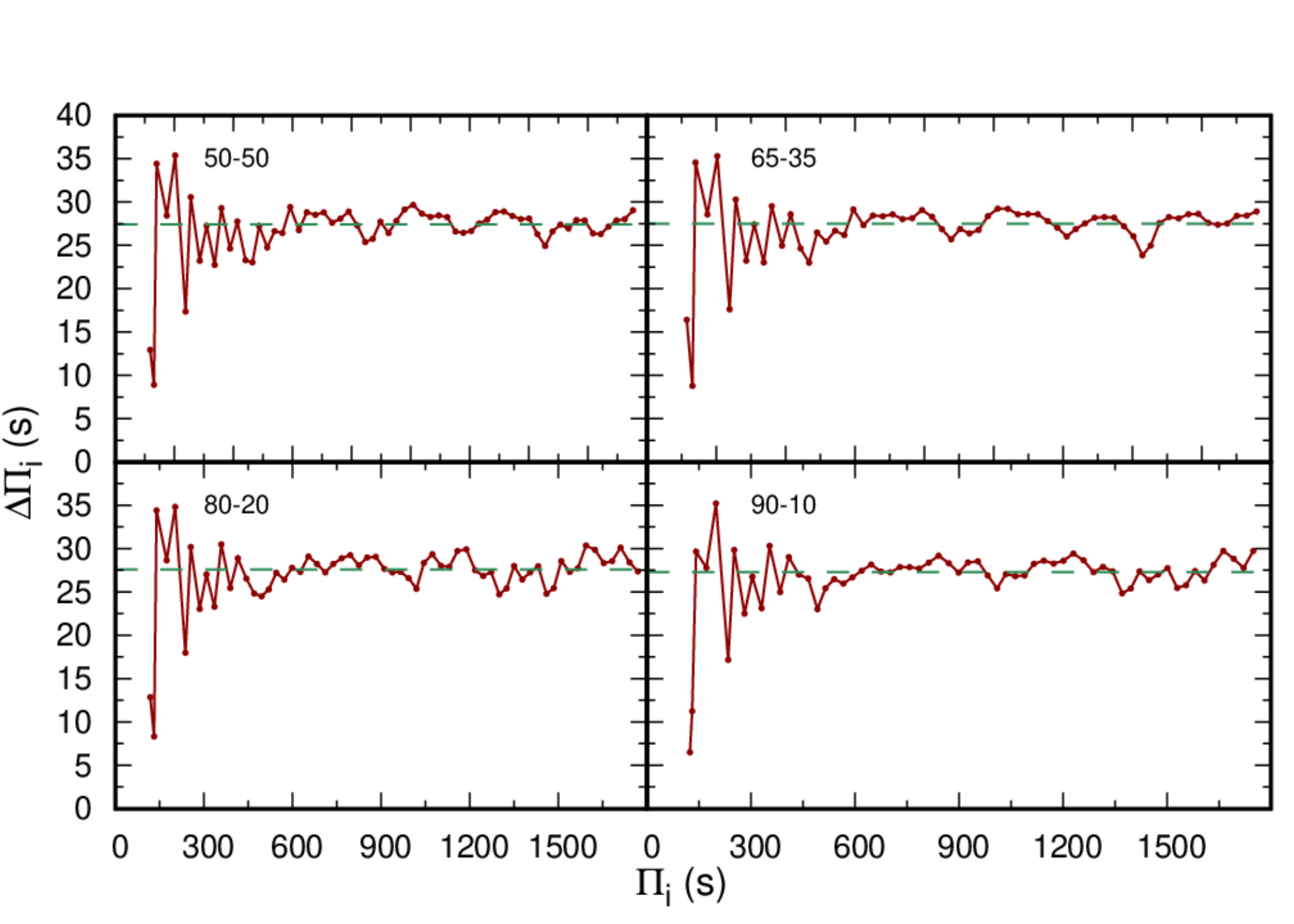}
   \caption{Same as Fig. \ref{fig:dp-rate}, but for the CF88 model computed with different branching ratios, as indicated in the upper left corner of each panel.}
  \label{fig:dp-br}
\end{figure}

    \begin{figure}
  \includegraphics[width=1\columnwidth]{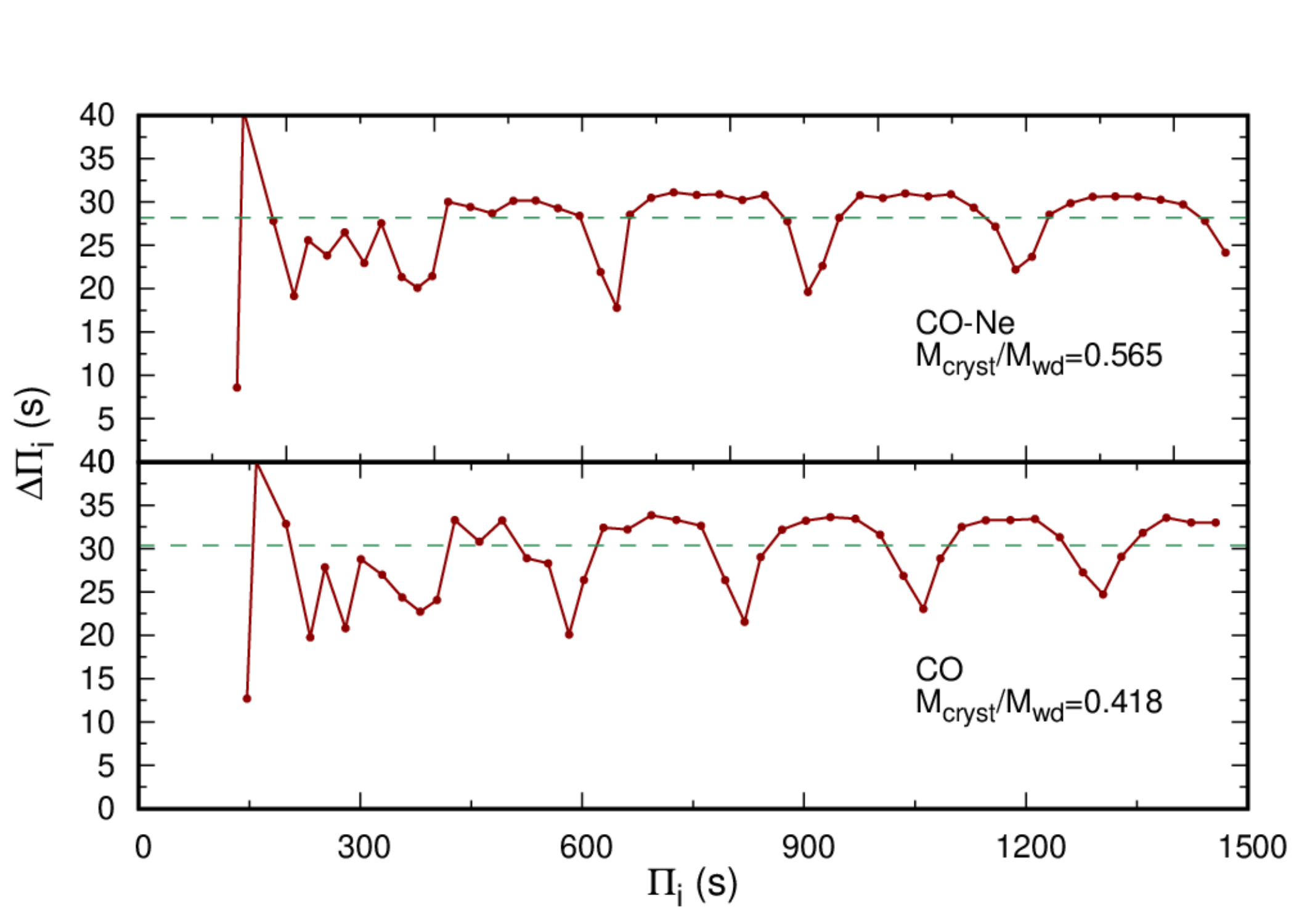}
    \caption{Same as Fig. \ref{fig:dp-rate}, but for the CO-core and hybrid WD models of Fig. \ref{fig:profile-685-12000} (upper and lower panels, respectively).}
   \label{fig:dp-CONe-CO}
\end{figure}

For the comparison of the pulsation properties, we computed the forward period spacing ($\Delta \Pi_k=\Pi_{k+1}-\Pi_k$), a quantity frequently used in asteroseismic analyses, that reflects the mode-trapping features of the models.
In Fig. \ref{fig:dp-rate}, we show the distribution of period spacing as a function of the periods for dipole modes, resulting from the assumption of different reaction rates. 
 As all the models have the same mass, the similarity in $\Delta \Pi_i$ reflects the fact that the crystallized portion of the star is similar in both cases, meaning that most of the different core-chemical features do not have a significant impact on the $g$-mode pulsations that might otherwise have allowed us to tell the different models apart.
 Consequently, the resulting asymptotic period spacing values are nearly identical.
The period spacing distributions are also very similar when considering
different branching ratios of the $^{12}{\rm C}+^{12}{\rm C}$ reaction rate during the
progenitor evolution, as shown by Fig. \ref{fig:dp-br}.

In contrast, the hybrid model shows perceptible differences in the trapping characteristics of the pulsation modes compared to the pure CO-core models, as seen in Fig. \ref{fig:dp-CONe-CO}.  
For $\Pi_i>300$~s, the CO-core model shows more frequent minima (a shorter trapping cycle) in their distribution of  $\Delta \Pi_i$ values than does the hybrid model. The reason for this is relatively straightforward. These trapped modes at long periods ($\Pi_i>500$~s) are modes with larger amplitudes in the homogeneous region above the crystallized core. As the outer border of the homogeneous region stays basically fixed, the larger the crystallized core, the smaller the homogeneous cavity in which the modes resonate. As shown by \cite{2019A&A...621A.100D}, this results in longer trapping cycles in that regime. It is worth noting that this difference in the trapping cycle of the two types of models would not be present if the mixed region were neutrally buoyant, as discussed by \cite{2024ApJ...961..197M}. The actual temperature profile above the crystallized core is a matter of ongoing discussion in the field \citep[e.g.][]{2024arXiv240201947C}. Additionally, differences in the asymptotic period amount to about $\sim 2$~s, this being a consequence of differences in the crystallization degree of the core.

\section{Conclusions}
\label{sect:concl}
In this work, we implemented the recently derived total nuclear reaction rate for carbon fusion, $^{12}$C+$^{12}$C, from \citet{2022A&A...660A..47M}, with the goal of evaluating its impact on computations of the structure and evolution of ultra-massive WDs and their progenitors, as compared to the case where the canonical $^{12}$C+$^{12}$C rate from CF88 is used. We also explored how the current uncertainty in the branching ratios for the $^{12}$C+$^{12}$C reaction's $\alpha$ and $p$ exit channels, which are the dominant ones at temperatures of astrophysical interest, affect the internal composition and evolution of these stars and their progenitors.

Our extensive numerical experiments were carried out using {\tt MESA}. Specifically, we computed evolutionary sequences, from the ZAMS up to the end of the C-burning phase, for models with initial masses $6.85\leq M_{\rm ZAMS}/M_{\odot}\leq8.50$ and a metallicity $Z=0.02$. In addition to exploring the $^{12}$C+$^{12}$C rates from \citet{2022A&A...660A..47M} and CF88, a wide range of $[\alpha/p]$ branching ratios was also considered, from  [50/50] up to [90/10]. The resulting structures were subsequently evolved, using {\tt LPCODE}, along the WD cooling track down to the ZZ Ceti (DAV) instability strip. When the models reached the latter phase, we computed their pulsation properties using the {\tt LP-PUL} code.


We found that using less efficient nuclear reaction rates results in a late onset of the C-burning phase, larger cores, higher burning temperatures, and, consequently, shorter C-burning phase lifetimes.
Despite these differences in the structure and evolution of SAGB progenitors, the impact on the distribution of the chemical elements is almost negligible. 
In contrast, the existing uncertainties in the relative efficiency of the $\alpha$ and $p$ exit channels constitute the most important uncertainty in determining the final chemical structures of SAGB progenitors.  Differences in the central $^{20}$Ne abundances can reach up to 17\% within the range of $[\alpha/p]$ ratios explored in our study, in the sense that a higher $^{20}$Ne content is achieved when the $\alpha$ channel is more dominant. Moreover, we found that higher production of $^{20}$Ne translates into smaller initial masses needed for C-ignition and longer C-burning phase lifetimes. 

An interesting result, derived from the exploration of the minimum mass needed for C-ignition, is that, for a specific range of masses that depend on the adopted total nuclear reaction rate and branching ratio (e.g., $6.85\leq M_{\rm ZAMS}/M_{\odot}<7.10$, for the CF88 rate and a branching ratio of [65/35]), carbon burns partially in the models' interiors. In such cases, the final chemical structure at the end of the C-burning phase consists of a CO-core surrounded by an ONe mantle. Depending on the amount of $^{20}$Ne produced during this stage, the progeny could be a CO-core WD or a hybrid CONe-core WD.

As for the impact on the ONe-core WD evolution due to the use of different nuclear reactions and branching ratios, we found differences in the cooling times and the size of the crystallized core of at most 1.3 and 0.8\%, respectively. We found that the impact on the pulsation properties of these stars is also negligible.

Regarding the hybrid CONe-core WDs, we found that they differ substantially in their evolution, compared with those composed of pure CO cores. Our results show that even as little as 10\% of $^{20}$Ne in its interior can modify the structure of the WD in such a way that crystallization starts earlier and, by the time the star reaches the ZZ Ceti instability strip, differences in the crystallized portion of the star and its cooling time can reach up to 15\% and 6\%, respectively.
This result is also reflected in the pulsation properties. By comparing its forward period spacing, we find that the hybrid  WD has less frequent minima (larger trapping cycle) than its CO-core counterpart. This is because the hybrid model has a larger crystallized core and, consequently, a smaller resonating cavity. 

In conclusion, our study reveals that current uncertainties in both the total $^{12}$C+$^{12}$C reaction rate {\em and} its branching ratios can have a significant impact upon the late stages of evolution of intermediate-mass stars and their progeny. As demonstrated by other authors, uncertainties in the $^{12}$C+$^{12}$C rate also impact the advanced evolutionary stages of higher-mass stars. Given its far-reaching astrophysical implications, further experimental work is thus urgently needed to properly constrain the $^{12}$C+$^{12}$C rate {\em and branching ratios} at astrophysically relevant energies.

\software{Modules for Experiments in Stellar Astrophysics
\citep[MESA;][]{Paxton2011, Paxton2013, Paxton2015, Paxton2018, Paxton2019}, {\tt MESASDK} 20.3.1 \citep{2020zndo...3706650T}, {\tt LPCODE}  \citep{2021A&A...646A..30A}, {\tt LP-PUL} \citep{2006A&A...454..863C}.}
\section*{acknowledgements} We wish to acknowledge the suggestions and comments
of the anonymous referee who strongly improved the original version of this
work. 
This work was supported by PIP 112-200801-00940 grant
from CONICET, grant G149 from the University of La Plata, PIP-2971 from CONICET (Argentina) and by
PICT 2020-03316 from Agencia I+D+i (Argentina). 
 This research has made use of the NASA Astrophysics Data System.

%
%


\bibliography{fran}
\bibliographystyle{aasjournal}







\end{document}